\documentclass{article}
\usepackage{amsmath}       
\usepackage{algorithm}     
\usepackage{algorithmic}   

\usepackage{float}
\usepackage[utf8]{inputenc} 
\usepackage[T1]{fontenc}    
\usepackage{hyperref}       
\usepackage{url}            
\usepackage{booktabs}       
\usepackage{amsfonts}       
\usepackage{nicefrac}       
\usepackage{microtype}      
\usepackage{lipsum}
\usepackage{graphicx}
\graphicspath{ {./images/} }

\title{DMCIE: Diffusion Model with Concatenation of Inputs and Errors to Improve the Accuracy of the Segmentation of Brain Tumors in MRI Images}

\author{
Sara Yavari\textsuperscript{1} \quad 
Rahul Nitin Pandya\textsuperscript{1} \quad 
Jacob Furst\textsuperscript{1} \\
\textsuperscript{1}School of Computing, DePaul University, Chicago, IL, USA \\
\texttt{\{syavari, npandya, jfurst\}@depaul.edu}
}

\begin{document}

\maketitle 
\begin{abstract}
Accurate segmentation of brain tumors in MRI scans is essential for reliable clinical diagnosis and effective treatment planning. Recently, diffusion models have demonstrated remarkable effectiveness in image generation and segmentation tasks. This paper introduces a novel approach to corrective segmentation based on diffusion models. We propose DMCIE (Diffusion Model with Concatenation of Inputs and Errors), a novel framework for accurate brain tumor segmentation in multi-modal MRI scans. We employ a 3D U-Net to generate an initial segmentation mask, from which an error map is generated by identifying the differences between the prediction and the ground truth. The error map, concatenated with the original MRI images, are used to guide a diffusion model. Using multimodal MRI inputs (T1, T1ce, T2, FLAIR), DMCIE effectively enhances segmentation accuracy by focusing on misclassified regions, guided by the original inputs. Evaluated on the BraTS2020 dataset, DMCIE outperforms several state-of-the-art diffusion-based segmentation methods, achieving a Dice Score of 93.46 and an HD95 of 5.94 mm. These results highlight the effectiveness of error-guided diffusion in producing precise and reliable brain tumor segmentations.

\textbf{keywords}: Brain Tumor Segmentation, Diffusion Model, 3D UNet, Error Reconstruction
\end{abstract}

\section{Introduction}
Medical image segmentation involves dividing a medical scan into distinct, meaningful regions, with numerous applications in healthcare, such as disease detection, treatment monitoring, surgical planning, and radiotherapy guidance. Precisely segmenting brain tumor areas in MRI scans is essential for successful tumor removal and radiotherapy, contributing significantly to improving diagnostic accuracy and optimizing therapeutic strategy. Despite significant advancements in medical image segmentation in recent years, various challenges remain \cite{Wang2022,Panayides2020} including the accurate segmentation of small or irregularly shaped tumors, which often leads to substantial errors\cite{Isensee2020}. In recent years, diffusion probabilistic models (DPMs) have garnered considerable attention as a potent category of generative models, particularly in the domain of computer vision \cite{Ho2020}. These models are highly proficient in producing high-fidelity images with outstanding diversity. Large-scale diffusion models, such as DALL-E2 \cite{Ramesh2022}, and Stable Diffusion \cite{Rombach2022}, have showcased remarkable capabilities in image synthesis \cite{Zhao2021,Goodfellow2020}. The diffusion mechanism operates as a structured Markov chain that emulates the stepwise introduction of noise into an image and subsequently learns to invert this process. It comprises two primary phases: the forward and reverse diffusion processes. In the forward phase, noise is systematically infused at each timestep, gradually converting the image into complete Gaussian noise. Conversely, the reverse phase starts with a Gaussian noise image and progressively restores the original image by systematically eliminating noise at each stage\cite{Ho2020}. This paradigm has emerged as one of the most prevalent approaches in computer vision, recognized for its efficiency in generating high quality images. 

The diffusion process demonstrates exceptional effectiveness across multiple domains, including image generation \cite{Rombach2022}, image editing \cite{Kawar2023}, and image segmentation \cite{Sun2023}, making it a versatile framework for various computer vision tasks, especially in medical imaging where it aids in tasks such as anomaly detection, tumor segmentation by generating high-quality synthetic data and improving model robustness \cite {Song2023}. 

\begin{figure}[H]
\centering
\includegraphics[width=0.8\textwidth]{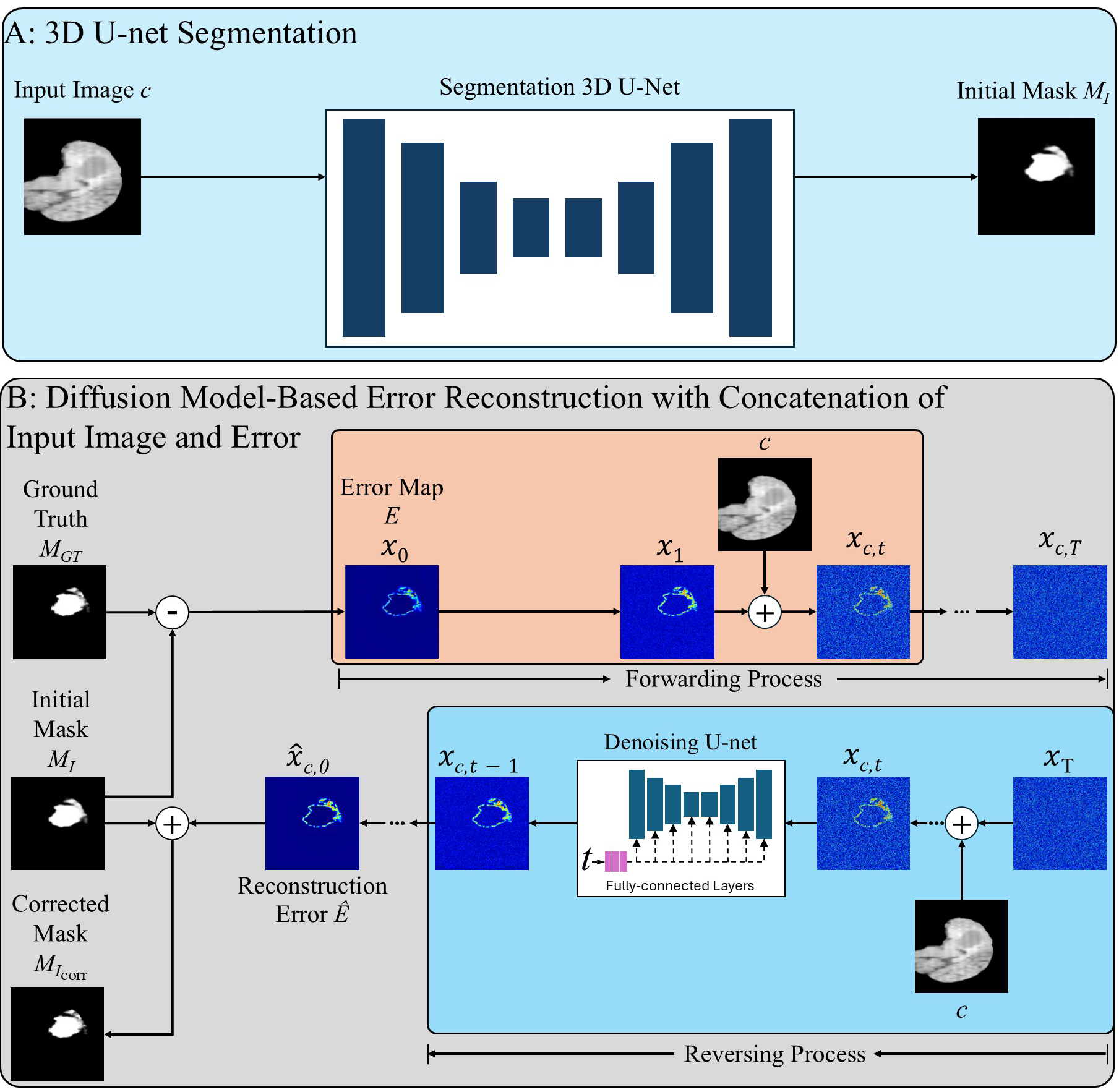}
\caption{
Overview of our proposed DMCIE framework for brain tumor segmentation using MRI. 
Part A shows how the input image $c$ is processed by a 3D U-Net to generate an initial segmentation mask $M_{I}$. 
Part B shows the diffusion model used to improve this initial mask. The error map $E$ is created by taking the difference between the ground truth mask $M$ and the initial mask $M_{I}$. In the forward process, noise is gradually added to the concatenation of the error map and the input image $c$ across multiple time steps. In the reverse process, the model removes the noise step by step. At each time step, the input image $c$ is concatenated with the noisy data to guide the model in predicting the corrected error. This helps the model focus more effectively on fixing segmentation mistakes. The final corrected error is added to the initial mask $M_{I}$ to produce the final improved segmentation result.}
\label{Idea2_7.jpg}
\end{figure}

In this work, we focus on binary brain tumor segmentation by merging all tumor subregions into a single foreground class, framing the task as a voxel-wise classification between tumor and non-tumor regions, assessing and localizing tumors. 
We propose DMCIE (Diffusion Model with Concatenation of Input and Error), a novel two-stage framework that refines initial segmentation masks through diffusion-based error correction, with its main contributions illustrated in Figure~\ref{Idea2_7.jpg}. as follows:

i) Initial Mask Generation: A 3D U-Net is used to produce an initial segmentation mask, as in Part A of Figure~\ref{Idea2_7.jpg}.

ii) Error Map Computation and Diffusion-Based Error Reconstruction: An error map is computed by the differences between the initial prediction from the ground truth masks. A diffusion model is then applied to reconstruct this error and refine the segmentation. During the diffusion process, the error map is concatenated with the original multi-modal MRI inputs (T1, T1ce, T2, and FLAIR) to guide the model in correcting misclassified regions. The reconstructed error is then integrated into the initial mask to generate the final corrected segmentation. We refer to this method as DMCIE to improve the segmentation accuracy of brain tumor segmentation based on MRI, as illustrated in Part B of Figure~\ref{Idea2_7.jpg}.

\section{Related work}
Magnetic Resonance Imaging (MRI) is widely recognized for its profound impact on radiology. MRI provides superior soft tissue contrast and supports multi-parametric, multi-axial, and multi-sequence imaging, making it an essential tool for neurological assessments \cite{re1}. Unlike Computed Tomography (CT), which relies on X-rays, MRI utilizes strong magnetic fields and radio- frequency pulses to generate detailed cross-sectional images of brain structures. Additionally, MRI enables multi-sequence acquisitions, including T1-weighted (T1w), T2-weighted (T2w), Fluid-Attenuated Inversion Recovery (FLAIR), and Diffusion-Weighted Imaging (DWI), all of which provide complementary information for tumor characterization and diagnosis \cite{re7}. Brain tumors are abnormal masses of tissue caused by uncontrolled cell proliferation in the brain. These tumors can be benign (non-cancerous) or malignant (cancerous), with the latter posing a significant risk due to aggressive invasion of surrounding tissue. Early detection and precise delineation of brain tumors are essential for accurate diagnosis, treatment planning, and prognosis. MRI-based tumor segmentation enables clinicians to assess tumor characteristics and develop personalized treatment strategies \cite{re3,re4}.

Traditional machine learning techniques for automated brain tumor segmentation in MRI images face challenges \cite{re5,re6}. Classic algorithms such as Support Vector Machines (SVMs), k-Nearest Neighbors (k-NN), and Decision Trees (DTs) have been employed for classification. While these techniques contributed to improving diagnostic objectivity, they were constrained by their reliance on manual feature selection and their inability to effectively process the high-dimensional and complex nature of MRI data \cite{re8}. Modern deep learning methods have significantly advanced MRI-based brain tumor segmentation with CNN-based architectures and attention mechanisms playing a crucial role in enhancing segmentation accuracy \cite{re9}. Despite these advancements, existing methods still face challenges in segmenting small, irregularly shaped tumors, resulting in segmentation errors. These difficulties arise due to limitations in handling complex variability in tumor shapes and appearances, the potential loss of spatial information during processing, and the need for more robust feature representation to generalize across diverse tumor morphologies  \cite{re8, re9}.

Various models have been designed to improve segmentation accuracy by incorporating advanced mechanisms. TransBTS \cite{re11} utilizes attention-based techniques to capture both fine-grained local features and broader contextual information, enhancing the precision of tumor boundary detection. Similarly, CANet \cite{re12} integrates a context-aware network, enabling a deeper understanding of intricate tumor structures and variations. Meanwhile, \cite{re13} introduced a segmentation method based on weighted patch extraction, which is integrated into the 3D U-Net architecture. This approach enhances segmentation accuracy by efficiently selecting and processing weighted patches, leading to more precise delineation of tumor interiors. Generative models offer an alternative approach to brain tumor segmentation by learning the underlying data distribution and generating new samples that preserve the structural properties of medical images. These models aim to learn the joint distribution between the input features \( x \) and the target labels \( y \), enabling them to estimate the conditional probability \( P(y \mid x) \). The maximum a posteriori probability (MAP) estimation is commonly employed to obtain segmentation predictions based on the learned distributions \cite{re1}. Variational Autoencoders (VAEs) \cite{re16} consist of two primary components: an encoder and a decoder, designed to extract meaningful data representations \cite{re17}. In the domain of brain tumor segmentation, VAEs play a crucial role in generating data from latent space while ensuring that the essential structural characteristics of the original medical images are preserved. Specifically, VAEs are employed for feature learning and dimensionality reduction in brain tumor segmentation tasks. The encoder transforms MRI brain images into a latent feature space, capturing key patterns and variations within the data. These compressed representations are then used to train segmentation models, improving feature extraction and enhancing the overall segmentation process\cite{re1}. 

\cite{re14} introduced a 3D MRI brain tumor segmentation method that integrates a VAE branch to regularize the encoder-decoder network, thereby improving segmentation performance. Similarly, \cite{re15} proposed a two-stage cascade model combining VAEs and attention gates, achieving notable accuracy in segmenting different tumor sub-regions. These approaches demonstrate that incorporating VAEs can effectively enhance the precision of brain tumor segmentation models. \cite{re18} proposed an unsupervised method that combines a dual autoencoder framework with Singular Value Decomposition (SVD) for feature optimization. This approach enhances the segmentation of brain tumors from MRI images by effectively capturing complex features and reducing dimensionality. \cite{re19} presented the Dual Residual Multi-Variational Auto-Encoder (DRM-VAE), a deep learning model designed to enhance brain tumor segmentation, particularly when certain MRI modalities are missing. This architecture integrates residual connections and multiple variational autoencoders (VAEs) to effectively merge available information, thereby improving segmentation accuracy. Generative Adversarial Networks (GANs), introduced by \cite{re20},are composed of two neural networks—a generator and a discriminator that engage in an adversarial process to improve data generation. The generator aims to synthesize realistic MRI images, while the discriminator evaluates their authenticity, distinguishing between real and generated samples. This adversarial training framework enables GANs to generate high-quality synthetic MRI images, which have been widely applied in medical imaging tasks, including data enhancement, anomaly detection, and segmentation enhancement\cite{re21}. Some studies have demonstrated that GAN-generated synthetic MRI scans can augment training datasets, addressing the issue of limited annotated data, thereby improving segmentation accuracy \cite{re21, re22}. Advanced GAN architectures, such as Vox2Vox, have been used for 3D brain tumor segmentation, achieving high Dice scores by effectively capturing spatial features \cite{re23}. 

Diffusion models have emerged as a powerful technique for brain tumor MRI segmentation\cite{re24,re25}.The stochastic nature of diffusion models introduces randomness during training, enhancing their flexibility and adaptability to complex tumor structures and morphological variations \cite{re25}. The study conducted by \cite{re26} proposed the use of Denoising Diffusion Probabilistic Models (DDPMs) for brain MRI segmentation, offering an innovative approach to DPM-based image segmentation. Their method generates synthetic labeled data, reducing the dependence on manually annotated pixel-wise segmentation. Although this technique represents a significant advancement, it remains computationally intensive and time-consuming. \cite{re27} PD-DDPM Model is a diffusion-based model designed for medical image segmentation, refining predictions through iterative denoising in the reverse diffusion process to improve accuracy and structural integrity. To enhance computational efficiency, it incorporates pre-segmentation results and leverages noise prediction based on forward diffusion rules, significantly accelerating segmentation while maintaining precision. 

ReCoSeg \cite{reCoSeg2025} was introduced as a residual-guided cross-modal diffusion framework that utilizes residual maps as attention cues between synthesized and original MRI modalities to improve segmentation accuracy with limited supervision. MedSegDiff \cite{re31} enhanced diffusion probabilistic models (DPMs) for medical image segmentation by incorporating a dynamic conditional encoding strategy, which reduces the adverse effects of high-frequency noise components using an FF resolver. To further improve the alignment between noise and semantic features, MedSegDiff-V2\cite{re32} was introduced, integrating a transformer-based architecture with Gaussian spatial attention blocks to enhance noise estimation and segmentation accuracy. The study \cite{re30} introduced SegDiff, a segmentation framework that integrates image data into the reverse diffusion process, enabling continuous and iterative refinement of segmentation results. This approach ensures that the segmentation process is progressively improved with each denoising step, enhancing accuracy and structural consistency. Our framework uses a diffusion-based segmentation approach that explicitly reconstructs segmentation errors to refine tumor boundaries. By concatenating error maps with multi-modal MRI inputs, our method guides the diffusion process to focus on misclassified regions. Unlike prior works that rely solely on generative noise prediction, we incorporate segmentation-aware supervision to directly optimize accuracy. Our approach aims to enhance binary tumor localization with improved boundary precision and clinical reliability.

\section{Methodology}
This section begins with a review of the 3D U-Net segmentation approach used for generating initial tumor masks, as outlined in Subsection~\ref{subsec:unet}. We then introduce our proposed framework, DMCIE, in Subsection~\ref{subsec:dmcie}, which refines these initial masks through diffusion-based error correction. This framework consists of two main components: the first~(\ref{subsubsec:error-map}) describes how the error map is generated to identify regions requiring refinement, while the second~(\ref{subsubsec:diffusion}) explains the diffusion process, including the forward and reverse steps, input concatenation, and the loss function used during training.

\subsection{3D U-Net Segmentation for Initial Mask Generation}
\label{subsec:unet}
In the first step (Part A), we use a 3D U-Net \cite{re28} as the initial tumor segmentation model to generate preliminary masks for the whole tumor (WT), the tumor core (TC) and the enhancement of the tumor (ET) \cite{re29} before refinement through diffusion-based reconstruction. In this stage, our primary objective is to obtain an initial segmentation mask for brain tumor images using the 3D U-Net segmentation model. Given an MRI image $c$ as input, Segmentation U-Net processes the image to produce the initial segmentation mask \( M_{I} \). Unlike 2D models, 3D U-Net processes volumetric MRI data, preserving spatial continuity across slices. Its encoder-decoder architecture with skip connections enables the retention of high-resolution spatial information while simultaneously extracting deep hierarchical tumor features \cite{re28}. These characteristics make the 3D U-Net an effective choice for accurately capturing the complex spatial structures of brain tumors in volumetric MRI scans. The 3D U-Net follows an encoder-decoder structure with skip connections, allowing it to retain high-resolution spatial information while extracting deep hierarchical tumor features. It consists of four resolution steps in both the analysis (encoder) and synthesis (decoder) paths. Each encoder block includes two 3×3×3 convolutions followed by ReLU activations and 2×2×2 max pooling for downsampling. In the decoder, each layer begins with a 2×2×2 transposed convolution with strides of 2 in all dimensions to upsample the feature maps, followed by two 3×3×3 convolutions each with ReLU activations. Skip connections are used between encoder and decoder layers at matching resolutions to preserve spatial detail. A final 1×1×1 convolution is used to reduce the output to the desired number of classes, followed by softmax activation for voxel-wise segmentation \cite{re28}.

\subsection{Diffusion Model-Based Error Reconstruction with Concatenation of Input MRI and Error Map (DMCIE)}
\label{subsec:dmcie}
After obtaining the initial segmentation from the 3D U-Net \cite{re28}, we introduce a diffusion model to refine the segmentation by reconstructing the error. In this process, the input MRI image is concatenated with the error map to improve the accuracy of segmentation.  This approach, referred to as DMCIE (Diffusion Model with Concatenation of Input and Error), is applied to MRI brain tumor segmentation. As illustrated in part B of Figure~\ref{Idea2_7.jpg}. This method simplifies and enhances the segmentation process by leveraging diffusion-based error correction.
 
\subsubsection{Generating the Error Map for Diffusion Model}
\label{subsubsec:error-map}
The primary reason for explicitly computing the error map before diffusion modeling is to focus refinement on regions that require correction rather than modifying the entire segmentation mask. This ensures that the diffusion model prioritizes misclassified tumor regions instead of over-modifying the mask and helps correct ambiguous tumor margins where the 3D U-Net struggles. The error map \( E \) is defined as the pixel-wise subtraction of the initial segmentation mask \( M_I \) and the ground truth segmentation mask \( M_{GT} \):

\begin{equation}
E = M_I - M_{GT}
\end{equation}

Here, \( M_I \) is the predicted tumor mask, and \( M_{GT} \) is the ground truth mask. 
 We denote the error map $E$ as \( x_0 \), which serves as the initial input for the diffusion process. The error map highlights the difference between the initial mask and the ground truth, providing the necessary input to start the diffusion process.

\subsubsection{Error Reconstruction via Diffusion Model and Input Image Concatenation}
\label{subsubsec:diffusion}
First, we prepare four grayscale multimodal MRI sequences T1, T2, T1ce, and T2-FLAIR each with a shape of (D, H, W), where D is the number of slices (depth), H is the height, and W is the width. Stacking these four modalities along the channel dimension results in a tensor of shape (4, D, H, W). Then error maps are normalized and concatenated with the four MRI modalities along the channel axis, forming a combined input tensor of shape (5, D, H, W). This structure enables the model to leverage both anatomical and modality-specific information from the MRI scans, as well as structural discrepancies highlighted by the error map. The model is better guided during training, allowing it to generate improved error maps and ultimately produce a more accurate segmentation mask.

The forward diffusion process is a fundamental component of the Diffusion Model \cite{Ho2020}. It is designed to gradually corrupt the error map \( x_0 \) by adding Gaussian noise at each timestep \( t \). The corrupted error map is then concatenated with the multi-modal MRI inputs (T1, T1ce, T2, FLAIR), as we can see in Figure~\ref{Idea2_7.jpg}, where \( x_0 \) is the initial error map, \( x_t \) represents the progressively corrupted error map at time step \( t \). \( x_{c,t} \) represents the concatenation of the multi-modal MRI inputs and the error map at step \( t \). \( x_{c,T} \) represents a pure Gaussian noise distribution after sufficient corruption steps. 
The goal of the model is to learn a reverse diffusion process that denoises \( x_{c,T} \) and reconstructs a more accurate error map by leveraging the multi-modal MRI inputs. The forward diffusion step follows a Markovian process\cite{Ho2020},
\begin{equation}
q(x_t \mid x_{t-1}) = \mathcal{N}(x_t; \sqrt{1 - \beta_t} \, x_{t-1}, \beta_t \mathbf{I})
\end{equation}

where \( \mathbf{I} \) represents the identity matrix, and \( \beta_1, \dots, \beta_T \) denote the variances of the forward diffusion process. At each step, a small amount of Gaussian noise is incrementally added to the image. Repeating this process for \( t \) steps, we can express it as:

\begin{equation}
q(x_t \mid x_0) := \mathcal{N}(x_t; \sqrt{\bar{\alpha}_t} x_0, (1 - \bar{\alpha}_t) \mathbf{I})
\end{equation}

where:
\begin{equation}
\alpha_t := 1 - \beta_t, \quad \bar{\alpha}_t := \prod_{s=1}^{t} \alpha_s
\end{equation}

We can directly express \( x_t \) as a function of \( x_0 \):

\begin{equation}
x_t = \sqrt{\bar{\alpha}_t} x_0 + \sqrt{1 - \bar{\alpha}_t} \, \epsilon, \quad \text{where } \epsilon \sim \mathcal{N}(0, \mathbf{I})
\end{equation}

In the reverse process\cite{Ho2020}, the model aims to reconstruct the original data by gradually removing noise from \( x_T \). The transition from \( x_t \) to \( x_{t-1} \) is learned using the model parameters \( \theta \) and is formulated as:

\begin{equation}
p_{\theta}(x_{t-1} \mid x_t) = \mathcal{N}(x_{t-1}; \mu_{\theta}(x_t, t), \Sigma_{\theta}(x_t, t))
\end{equation}

As demonstrated in \cite{Ho2020}, we can estimate \( x_{t-1} \) from \( x_t \) using the following equation:

\begin{equation}
x_{t-1} = \frac{1}{\sqrt{\alpha_t}} \left( x_t - \frac{1 - \alpha_t}{\sqrt{1 - \bar{\alpha}_t}} \, \epsilon_{\theta}(x_t, t) \right) + \sigma_t z, \quad \text{where } z \sim \mathcal{N}(0, \mathbf{I})
\label{eq:reverse_xt}
\end{equation}

Here, \( \sigma_t \) represents the variance schedule, which can be either fixed or learned by the model \cite{nic}. The term \( \epsilon_{\theta}(x_t, t) \) is estimated using a deep neural network (typically a U-Net), which predicts the noise component that needs to be removed at each step.

By iteratively applying this denoising process, the model progressively reconstructs the original error map from the fully corrupted \( x_{c,T} \). This refined reconstruction is essential for enhancing segmentation performance, as it leverages the structured noise removal process inherent to diffusion models.

The reverse process \( p_\theta \) is learned by the model parameters \( \theta \) and is defined as:

\begin{equation}
p_\theta(x_{t-1} \mid x_t) := \mathcal{N}(x_{t-1}; \mu_\theta(x_t, t), \Sigma_\theta(x_t, t))
\end{equation}

Following the work of \cite{Ho2020}, the reverse diffusion process aims to reconstruct a cleaner version of \( x_t \) at each step by gradually removing noise. 

\begin{equation}
z \sim \mathcal{N}(0, \mathbf{I})
\end{equation}

is a random Gaussian noise component, which introduces stochasticity into the sampling process. 

The given equation describes the transition from \( x_t \) to \( x_{t-1} \):

\begin{equation}
x_{t-1} = \frac{1}{\sqrt{\alpha_t}} \left( x_t - \frac{1 - \alpha_t}{\sqrt{1 - \bar{\alpha}_t}} \, \epsilon_{\theta}(x_t, t) \right) + \sigma_t z
\end{equation}

where \( \alpha_t \) is the noise scaling factor, \( \bar{\alpha}_t \) is the cumulative noise product, \( \epsilon_{\theta}(x_t, t) \) is the predicted noise, and  \( \sigma_t \) represents the variance schedule, which can be learned by the model, as introduced by \cite{nic}. We leverage diffusion models with multi-modal MRI images, denoted as \( c \), to enhance semantic segmentation. Here, \( {\text{T1}}, {\text{T1ce}}, {\text{T2}}, {\text{FLAIR}} \) represent different MRI sequences, each capturing unique tissue contrasts, collectively referred to as \( c \). Additionally, \( x_1 \) represents a noisy version of the initial error map \( x_0 \) during the diffusion process. The concatenation process thus can be 

\begin{equation}
x_{\text{c,t}} = \text{concat}(c, x_{\text{1}})
\end{equation}

We have:
\begin{equation}
x_{c,t-1} = \frac{1}{\sqrt{\alpha_t}} \left( x_{c,t} - \frac{1 - \alpha_t}{\sqrt{1 - \bar{\alpha}_t}} \, \epsilon_{\theta}(x_{c,t}, t) \right) + \sigma_t z, \quad \text{where } z \sim \mathcal{N}(0, \mathbf{I})
\label{eq:concat_reverse}
\end{equation}

The loss function used to train the diffusion model is defined as:

\begin{equation}
\mathcal{L}_{\text{ConcatDiff}} = \mathbb{E}_{\mathbf{x}_0, \epsilon, t} \left[ 
\text{BCE}\left( \epsilon, \epsilon_\theta(\mathbf{x}_{c,t}, t) \right) + 
\lambda \cdot \text{DiceLoss}\left( \epsilon, \epsilon_\theta(\mathbf{x}_{c,t}, t) \right)
\right]
\label{eq:concatdiff}
\end{equation}

The loss function $\mathcal{L}_{\text{ConcatDiff}}$ combines Binary Cross-Entropy (BCE) and Dice loss, both applied to the predicted noise $\epsilon_\theta(x_{c,t}, t)$ at each diffusion time step $t$, along with the original input $x_0$, the true noise $\epsilon$, and the time step $t$. Here, $x_{c,t}$ denotes the concatenation of the condition $c$ and the noisy input at time step $t$. BCE measures the pixel-wise differences between the predicted and actual noise, while Dice loss emphasizes the overlap between them, effectively addressing class imbalance in segmentation tasks. $\lambda$ is a hyperparameter that balances the contribution of Dice loss relative to BCE and is typically tuned based on validation performance.
In the proposed DMCIE framework, the reconstructed error map \( \hat{E} \) is added to the initial mask \( M_I \) to produce the final corrected mask, improving the accuracy of brain tumor segmentation. The corrected mask is computed as:

\begin{equation}
M_{I_{\text{corr}}} = M_I + \hat{E}
\end{equation}

\section{Experimental Setup}

\subsection{Dataset}
We evaluated our approach using the BraTS2020 dataset, which provides multi-modal MRI scans for 355 subjects, each including four MRI sequences: T1, T1ce, T2, and FLAIR. The four MRI sequences are stacked to create a 4-channel 3D input for each subject. For preprocessing, we discard the uppermost 26 and lowermost 80 axial slices from each 3D volume, where tumors rarely appear, resulting in 78 informative slices per subject. Each volume is then intensity-normalized by clipping the top and bottom 1st percentile of voxel intensities, followed by center cropping and resizing to a uniform shape of (4, 78, 120, 120). The ground truth annotations consist of three tumor subregions: Whole Tumor (WT), Tumor Core (TC), and Enhancing Tumor (ET). These are merged into a single foreground class to frame the task as binary tumor segmentation, where each voxel is classified as either tumor or non-tumor.

\subsection {Implementation Details}



In our implementation, during training, the input to the diffusion model includes the four MRI modalities and the error map, resulting in a 5-channel volume of shape $(5, 78, 120, 120)$, where $D = 78$, $H = 120$, and $W = 120$. This configuration enables the model to jointly consider both anatomical information from the MRI images and spatial discrepancies captured by the error map. As a result, the model is better guided during training, leading to the generation of improved error maps and more accurate segmentation masks. The 3D U-Net is first trained using a hybrid Dice + Binary Cross-Entropy loss to predict initial tumor masks. Subsequently, the Denoising Diffusion Probabilistic Model (DDPM) is trained using the $\mathcal{L}_{\text{ConcatDiff}}$ loss (see Eq.~\ref{eq:concatdiff}), which combines BCE and Dice loss for noise prediction. The DDPM follows a noise schedule over $T = 1000$ timesteps, ensuring smooth and progressive refinement of segmentation errors. Optimization is performed using the Adam optimizer with a learning rate of $7 \times 10^{-4}$ for the U-Net and $3 \times 10^{-4}$ (with $1 \times 10^{-5}$ weight decay) for the diffusion model. We train the models on a GPU-enabled system equipped with a NVIDIA RTX 3090 with 24GB VRAM. The CPU-based inference is tested on an Intel Core i9-12900K processor with 64GB RAM, ensuring that the trained model is optimized for deployment across different hardware configurations.

\section {Evaluation Metrics and Experiment Results}

We employ two widely adopted evaluation metrics in medical image analysis to assess the performance of our brain tumor segmentation model: the Dice Similarity Coefficient (Dice Score), which quantifies the volumetric overlap between the predicted segmentation and the ground truth, and the 95th percentile Hausdorff Distance (HD95), which captures the boundary-level discrepancy -crucial for evaluating segmentation accuracy in clinical contexts. We conducted extensive experiments on the BraTS2020 dataset, focusing on binary segmentation of the Whole Tumor (WT) region. To enable a fair and reproducible comparison, we reimplemented four recent diffusion-based tumor segmentation models CorrDiff ~\cite{corrdiff2023}, SF-Diff   ~\cite{sfdiff2024}, MedSegDiff ~\cite{medsegdiff2022}, and BerDiff ~\cite{berdiff2023} -within a unified training and evaluation pipeline. All models were trained using identical data splits, preprocessing strategies, and evaluation protocols on the BraTS2020 dataset for binary segmentation. CorrDiff was implemented with a UNet++ encoder and a DDPM-based corrective decoder incorporating class tokens. Although this model originally designed for multi-class tumor segmentation, we adapted CorrDiff to focus on WT-only masks. SF-Diff followed a lightweight DDPM design tailored for binary segmentation and employed timestep conditioning with a binary loss. MedSegDiff utilized a UNet3+ backbone enriched with frequency-aware modules and dynamic conditioning for suppressing irrelevant high-frequency signals. BerDiff, on the other hand, adopted a Bernoulli sampling mechanism with XOR-based stochastic corruption and calibration layers to simulate binary mask uncertainty. All models were trained using a consistent linear noise schedule with 1,000 timesteps, identical optimization strategies, and were evaluated using the same loss function -a combination of Binary Cross-Entropy (BCE) and Dice Loss -along with uniform stopping criteria, to ensure that any performance variations arise solely from architectural differences. As shown in Table~\ref{tab:brats_diffusion_comparison}, our proposed DMCIE achieved a Dice Score of 93.46\% and an HD95 of 5.94 mm, outperforming all reimplemented baselines across both metrics.

\begin{table}[h]
\centering
\caption{Quantitative comparison of Dice and HD95 on BraTS 2020}
\begin{tabular}{lcc}
\toprule
\textbf{Model} & \textbf{Dice Score (\%)} $\uparrow$ & \textbf{HD95 (mm)} $\downarrow$ \\
\midrule
CorrDiff ~\cite{corrdiff2023} & 90.68 & 7.88 \\
SF-Diff ~\cite{sfdiff2024} & 92.03 & 6.82 \\
MedSegDiff ~\cite{medsegdiff2022} & 91.32 & 8.68 \\
BerDiff ~\cite{berdiff2023} & 89.98 & 8.56 \\
\textbf{Ours (DMCIE)} & \textbf{93.46} & \textbf{5.94} \\
\bottomrule
\end{tabular}
\label{tab:brats_diffusion_comparison}
\end{table}


\begin{figure}[h]
    \centering
    \includegraphics[width=0.6\textwidth]{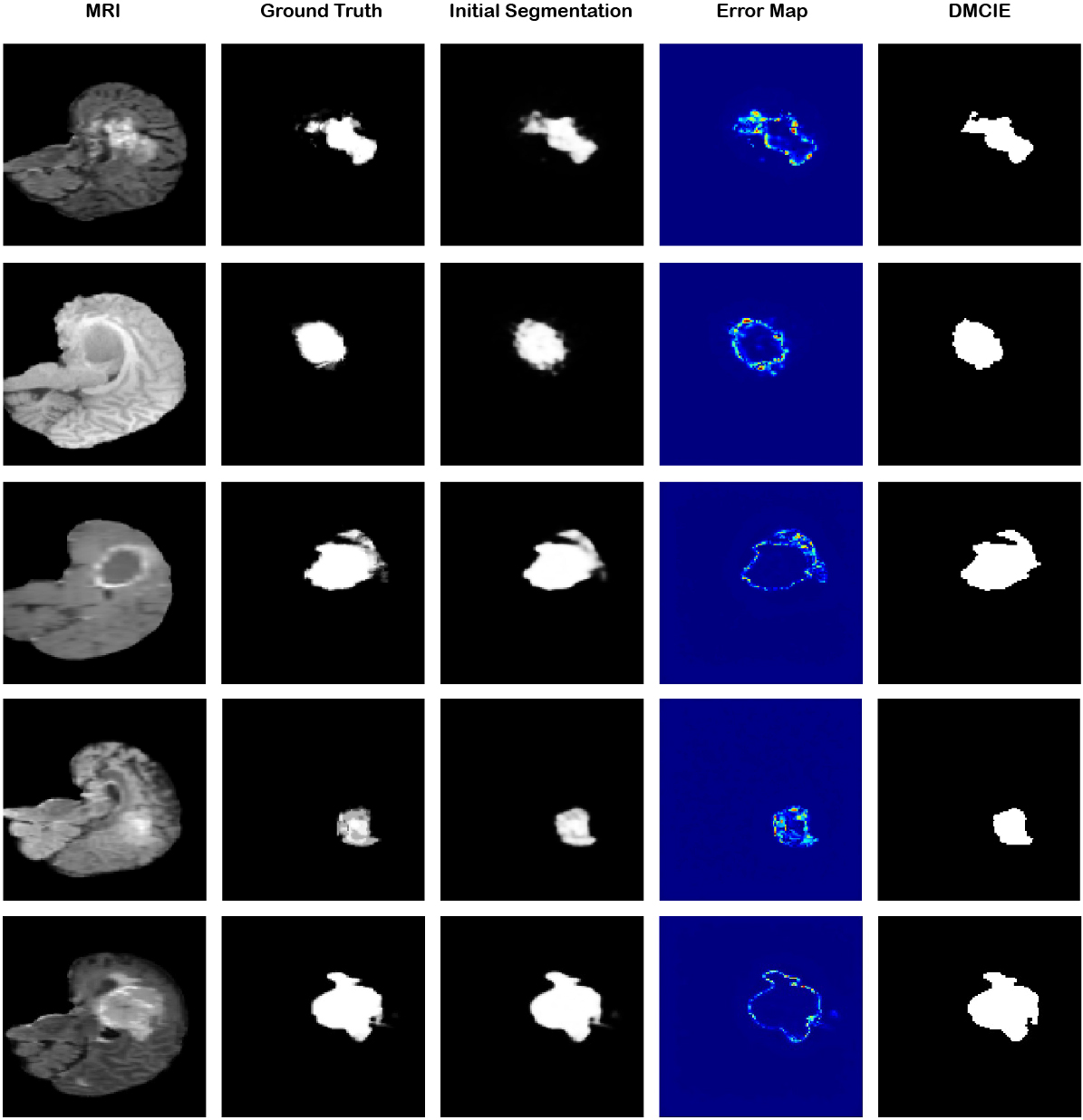}
    \caption{Visual illustration of the DMCIE segmentation pipeline using data from the BraTS2020 dataset. Columns represent the input MRI, ground truth, initial U-Net prediction, generated error map, and the final refined output from DMCIE. The error map highlights regions of disagreement and guides the diffusion model to refine boundaries and correct segmentation errors, resulting in improved alignment with the ground truth.}
    \label{fig:figure1}
\end{figure}

\vspace{0.5cm}

\begin{figure}[H]
    \centering
    \includegraphics[width=0.9\textwidth]{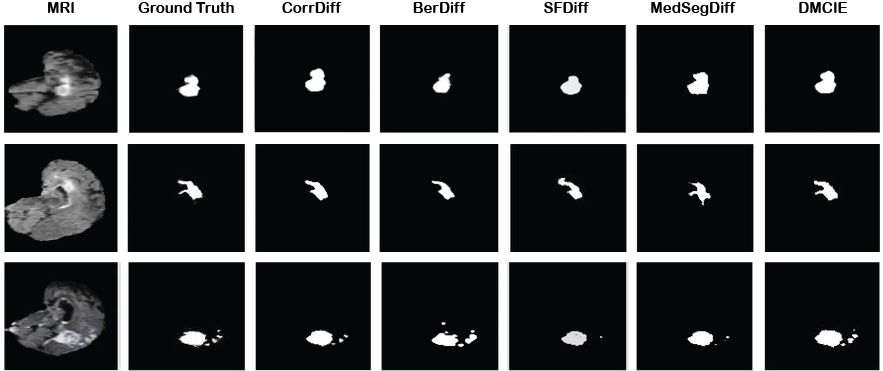}
    \caption{Illustration of the DMCIE segmentation pipeline on BraTS2020 samples. Columns show the input MRI, ground truth, initial U-Net prediction, generated error map, and the final output produced by DMCIE. The error map highlights uncertain regions and guides the diffusion model to refine boundaries and correct segmentation errors, resulting in improved alignment with the ground truth.}
    \label{fig:figure2}
\end{figure}

As shown in Figure~\ref{fig:figure1},the internal workflow of the DMCIE model, showcasing five representative test cases from the BraTS2020 dataset. For each example, we display the input MRI (either T2 or FLAIR), the ground truth tumor mask, the initial prediction generated by a 3D U-Net, the computed error map, and the final output produced by DMCIE. The error map functions as an attention mechanism, highlighting regions of disagreement between the initial segmentation and the ground truth. These error-prone areas -typically located along tumor boundaries- are then refined through the reverse diffusion process. This leads to segmentation outputs that are not only more accurate but also better aligned with the ground truth, especially in challenging cases involving fine structural irregularities or incomplete contours. Figure~\ref{fig:figure2} presents a qualitative comparison of segmentation results between DMCIE and other state-of-the-art methods on the BraTS 2020 dataset. Each row corresponds to a distinct test case, displaying the input MRI, ground truth, and predictions from CorrDiff ~\cite{corrdiff2023}, BerDiff ~\cite{berdiff2023}, SF-Diff   ~\cite{sfdiff2024}, MedSegDiff ~\cite{medsegdiff2022}, and our proposed DMCIE model. Our model achieved superior results by generating enhanced error maps through concatenation with input images during the diffusion process. Furthermore, the use of multi-modal MRI input improves anatomical context awareness, allowing DMCIE to generalize more effectively across diverse tumor morphologies

\subsection {Conclusion}
In this study, we proposed DMCIE, a diffusion-based segmentation framework that integrates error-guided correction and multi-modal MRI context to enhance binary brain tumor segmentation. Unlike prior models that rely solely on generative reconstruction, DMCIE leverages the structured difference between initial predictions and ground truth masks to explicitly guide the diffusion process. This design allows the model to concentrate on refining error-prone regions, particularly along complex tumor boundaries. Through extensive evaluation on the BraTS2020 dataset, DMCIE consistently outperformed recent diffusion-based baselines, achieving state-of-the-art results in both Dice Score and HD95. Our results demonstrate that incorporating segmentation-aware supervision and explicit error modeling into the diffusion framework leads to more accurate, and anatomically precise tumor segmentations. Given its strong performance and architectural flexibility, DMCIE offers a promising direction for deploying diffusion models in real-world clinical workflows for neuro-oncology.

\bibliographystyle{unsrt}  

\bibliography{Ref}     
\end{document}